\documentclass[amssymb,useAMS,prd,aps,twocolumn,superscriptaddress,preprintnumbers,nofootinbib]{revtex4}
\usepackage{pslatex}
\usepackage[pdftex]{graphicx}
\usepackage{psfrag}
\usepackage{epsfig}
\usepackage{color}
\usepackage{cancel}

\def\refe@jnl#1{{#1}}
\def\aj{\refe@jnl{Astron.~J.}}
\def\araa{\refe@jnl{Annu.~Rev.~Astron.~Astrophys.}}
\def\apj{\refe@jnl{Astrophys.~J.}}
\def\apjl{\refe@jnl{Astrophys.~J.~Lett.}}
\def\aap{\refe@jnl{Astron.~Astrophys.}}
\def\mnras{\refe@jnl{Mon.~Not.~R.~Astron.~Soc.}}
\def\prd{\refe@jnl{Phys.~Rev.~D}}
\def\fcp{\refe@jnl{Fund.~Cos.~Phys.}}
\def\physrep{\refe@jnl{Phys.~Rep.}}
\def\physlett{\refe@jnl{Phys.~Lett.}}

\def\invisible#1{  }

\def\lsim{\mathrel{\lower4pt\hbox{$\sim$}} 
\hskip-9.5pt\raise1.6pt\hbox{$<$}\;} 
 
\def\gsim{\mathrel{\lower4pt\hbox{$\sim$}} 
\hskip-9.5pt\raise1.6pt\hbox{$>$}\;}

\begin{document}
\preprint{ULB-TH/12-11}

\title{What if Dark Matter Gamma-Ray Lines come with Gluon Lines?}

\author{Xiaoyong, Chu}
\affiliation{Service de Physique Th\'eorique\\
 Universit\'e Libre de Bruxelles\\ 
Boulevard du Triomphe, CP225, 1050 Brussels, Belgium}
\author{Thomas Hambye}
\affiliation{Service de Physique Th\'eorique\\
 Universit\'e Libre de Bruxelles\\ 
Boulevard du Triomphe, CP225, 1050 Brussels, Belgium}
\affiliation{Departamento de F\'isica Te\'orica, Universidad Aut\'onoma de Madrid and\\ 
Instituto de F\'isica Te\'orica IFT-UAM/CSIC, Cantoblanco, 28049 Madrid, Spain }
\author{Tiziana Scarna} 
\author{Michel H.G. Tytgat}
\email[]{xiaochu@ulb.ac.be;thambye@ulb.ac.be;tscarna@ulb.ac.be;mtytgat@ulb.ac.be}
\affiliation{Service de Physique Th\'eorique\\
 Universit\'e Libre de Bruxelles\\ 
Boulevard du Triomphe, CP225, 1050 Brussels, Belgium}


\begin{abstract}

In dark matter (DM) models, the production of a $\gamma$ line (or of a ``box-shaped'' $\gamma$-ray spectrum) from  DM annihilation proceeds in general from a loop diagram involving a heavy charged particle. If the charged particle in the loop carries also a color charge, this leads inevitably to DM annihilation to gluons, with a naturally larger rate. We consider a scenario in which DM candidates annihilate dominantly into gluon pairs, and determine (as far as possible, model-independent) constraints from a variety of observables: a) the dark matter relic density, b) the production of anti-protons, c) DM direct detection and d) gluon-gluon fusion processes at LHC.
Among other things, we show that this scenario together with the recent claim for a possible $\gamma$ line from the Galactic center
in the Fermi-LAT data, leads to a relic abundance of DM that may be naturally close to the cosmological observations. 


\end{abstract}

 \maketitle
\section{Introduction}
The trouble with indirect searches for dark matter is that in general the possible signatures, like  gamma rays and antimatter in cosmic rays, may not be easily discriminated from those of more mundane, astrophysical sources. Two notable exceptions are the {possibilities} that dark matter may produce significant gamma-ray monochromatic lines \cite{Bergstrom:1988fp}, or that it may be captured by the Sun  \cite{Silk:1985ax} or the Earth \cite{Freese:1985qw}, where it may produce high-energy neutrinos. These ``smoking guns'' for dark matter, which have no expected astrophysical counterparts, are actively sought for by various experiments but with, so far, no conclusive result. In particular the Fermi-LAT collaboration has put limits on the possible monochromatic flux \cite{Ackermann:2012qk} from dark matter annihilation or decay (see also \cite{Vertongen:2011mu}). Recently, it has been claimed that the current Fermi-LAT data may actually contain some indication of a monochromatic gamma-ray signal in the vicinity of the Galactic centre, with $E_\gamma \sim 130$ GeV \cite{Weniger:2012tx}. This exciting, albeit tentative result, has been already challenged in \cite{Profumo:2012tr}, where it is claimed that the excess may have an astrophysical explanation, possibly related to the so-called Fermi bubbles regions (see also \cite{Boyarsky:2012ca}). Some support for a monochromatic line is given in \cite{Tempel:2012ey,Su:2012}, in which it is furthermore claimed that the signal is actually consistent with dominant emission from the Galactic centre. This exciting possibility has already attracted some attention in the recent literature \cite{Chalons:2012hf,Ibarra:2012dw,Dudas:2012pb,Cline:2012nw,Choi:2012ap,Kyae:2012vi,Lee:2012bq,Rajaraman:2012db,Acharya:2012dz,Garny:2012vt,Buckley:2012ws}.

\begin{figure}[!htb]
\centering
\includegraphics[height=2cm]{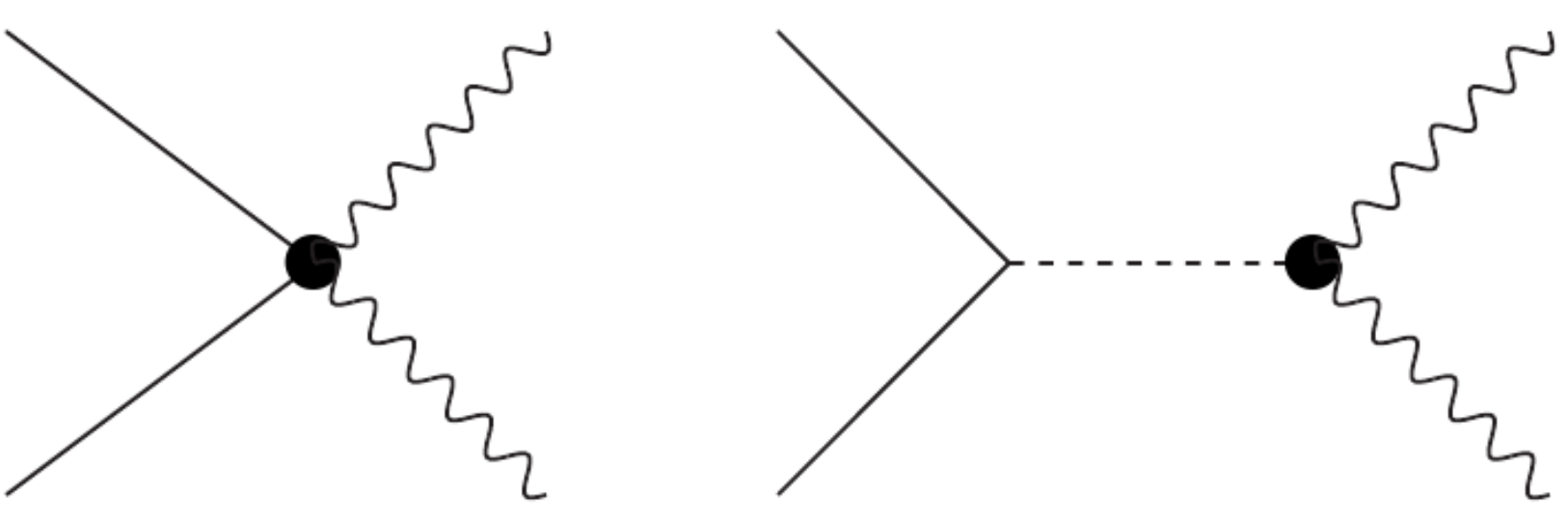}
\caption{Topologies with pairs of gamma-ray lines. The black dot is meant to represent a one-loop induced coupling.}
\label{fig:2gammatopologies}
\end{figure}

To produce a $\gamma$-ray spectrum from DM annihilation with sharp features, such as a $\gamma$ line (or more generally a narrow box-shaped spectrum \cite{Ibarra:2012dw}), in general one needs to invoke a Feynman diagram with a charged intermediate particle (see {\em e.g.}~\cite{Gustafsson:2007pc,Jackson:2009kg,Arina:2009uq} for non-SUSY instances), which can manifest itself only at the one-loop level (an exception to this rule of thumb is the scenario of Ref.\cite{Mambrini:2009ad}). If we limit ourselves to one-loop induced couplings, then there are two topologies for a $\gamma$ line, see Fig.\ref{fig:2gammatopologies}: either through a quartic vertex, or through an s-channel annihilation into an intermediate particle $S$ that couples to two photons. The former is an effective operator of dimension $7$ ($6$) for a fermionic (scalar) DM candidate. While  the two processes  become indistinguishable in the limit $m_S \gg 2 m_{DM}$, the quartic coupling does not necessarily require a UV completion with an intermediate particle: a box diagram for instance, like in \cite{Cline:2012nw}, or in the MSSM \cite{Rudaz:1986db,Drees:1993bh}, is another common possibility. 

\begin{figure}[!htb]
\centering
\includegraphics[height=2.5cm]{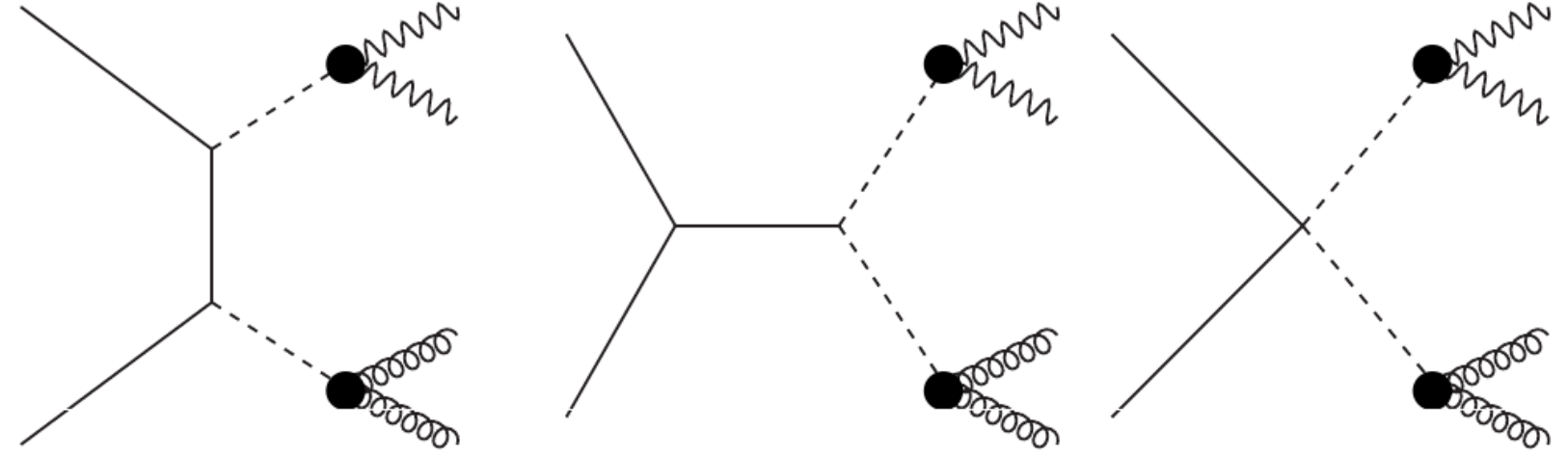}
\caption{Topologies giving to a box-shaped gamma spectrum. Annihilation proceeds through the decay of a, on-mass-shell, messenger particle. A larger coupling to gluons implies that the dominant contribution comes from decay of the messengers into two gammas pairs and two gluons pairs. }
\label{fig:boxtopologies}
\end{figure}

Similarly for a box-shaped spectrum, were the particle(s) that produce the $\gamma$ are on-shell,  there are three topologies (Fig.\ref{fig:boxtopologies}) still at one-loop: $t$, $s$ or quartic channel annihilation of DM to two on-shell particles, or messengers, that we call generically $S$ (there may be more than one $S$), with subsequent one-loop two-body decay of $S$. The box-shaped scenario, unlike the $\gamma$ line scenario, has the particularity that the production of hard $\gamma$ is not necessarily one-loop suppressed, as it is proportional to the tree-level production of on-shell $S$ particles, and to their branching ratio into photons (this is a particular instance of secluded DM \cite{Pospelov:2007mp}).
This allows  to produce a large flux of $\gamma$, even if the couplings are small. 
Familiar examples of $S$ neutral particle decaying to photons are 
scalar fields ($\pi^0$, axions, Higgs) which 
decay into $\gamma$ through a loop diagram with charged particles.

In the following we analyse the consequences of a scenario where some of the charged particles in the loop  carry color,
 in which case one has inevitably similar diagrams with a pair of gluons instead of photons.
One important property of these one-loop diagrams is that the ratio of gluon-to-photon production has little dependence on the topology one considers in Fig.~1. 
Another one is that the production of gluons is naturally larger than the production of photons. As an illustration, let us consider the example of coupling to gluons and photons pairs through a one-loop, triangular diagram 
with a single heavy vector quark of electric charge $Q_e$, and no weak isospin.
In this case the ratio of gluons to prompt gammas produced per annihilation is
\begin{equation}
\label{eq:ratio}
C = {\langle \sigma v\rangle_{gg}\over \langle \sigma v\rangle_{\gamma\gamma}} \qquad \mbox{\rm (line spectrum)}
\label{Rggammaline}
\end{equation}
with $C$ a numerical constant that, in the heavy quark limit ($m_Q \gg m_{DM}$), is \footnote{
Notice  that, if the ${\cal O}(\alpha_S)$ corrections to  $\gamma\gamma$ production are small, there are  substantial for decay into  gluons,  multiplying the ratio of Eq.(\ref{eq:goldenratio}) by a factor of about 2 (see {\em e.g.}~\cite{Djouadi:2005gi}). Unless stated, we do not consider explicitly these NLO corrections, as our approach is essentially model independent. In other words, such corrections are included in the definition of $C$. }
\begin{equation}
\label{eq:goldenratio}
C = { 2\alpha_S^2\over 9 Q_e^4 \alpha^2} \simeq 44.
\end{equation}
For the numerical estimate we have set $Q_e=1$ and have we taken $\alpha_S$ and $\alpha$ at  the scale $m_{DM} =  130$ GeV. 

In the second equality of (\ref{Rggammaline}) we have not taken into account the contribution from the $\langle \sigma v\rangle_{\gamma Z}$ and $\langle \sigma v\rangle_{ZZ}$ channels. These are automatically present, but they are sub-leading and we neglect them for the sake of our argument. Now, {if there are no other  annihilation channels than the one-loop processes discussed so far}, then the total annihilation cross section may be approximated by
\begin{equation}
{\langle \sigma v\rangle_{tot}\over\langle \sigma v\rangle_{\gamma\gamma}}\simeq 1+C \approx C\qquad \mbox{\rm (line spectrum)}
\label{ratiosigmaline}
\end{equation}
so that, unless the electric charge is larger than $Q_e=1$, one gets naturally a large ratio. 

Similarly, in the case of  a narrow box-shaped spectrum, one gets (neglecting again the sub-leading decay of  $S$ to $\gamma Z$ and $ZZ$):
\begin{equation}
C =  \frac{\Gamma(S \rightarrow g g)}{\Gamma(S\rightarrow \gamma\gamma)}\qquad \mbox{\rm (box-shaped spectrum)}
\end{equation}
just as in (\ref{eq:ratio}). 
Taking into account that $C$ is naturally much larger than one, most of the photons are produced in the process DM DM$\rightarrow gg + \gamma\gamma$, which implies that
\begin{equation}
{\langle \sigma v\rangle_{tot}\over\langle \sigma v\rangle_{\gamma\gamma}}\simeq 1+{C\over 2}\approx {C\over 2}\qquad \mbox{\rm (box-shaped spectrum)}
\label{ratiosigmabox}
\end{equation}

The results of Eq.~(\ref{ratiosigmaline}) and Eq.~(\ref{ratiosigmabox}) are interesting in two ways. First of all, one can compare these ratios with the ratio of the canonical cross section required for thermal freeze-out, 
\begin{equation}
\langle \sigma v\rangle_{FO}\approx 3\cdot 10^{-26} \, cm^2\cdot s^{-1}
\label{sigmafo}
\end{equation}
to the $\gamma\gamma$  cross section for the 130 GeV line claimed
in Ref.\cite{Weniger:2012tx} (for an Einasto profile),
\begin{equation}
\label{eq:sigammagammabis}
\langle \sigma v\rangle_{\gamma\gamma} \approx 1.2 \cdot 10^{-27}\, cm^2\cdot s^{-1},
\end{equation}
This ratio equals  $\sim25$, very close to the values quoted above for $Q_e=1$!
Therefore, the  scenario we consider may give a natural explanation for the strength of such a $\gamma$ line. 
Second, and this is of broader scope, it is important to confront such a scenario to other indirect, direct and collider constraints. In particular, the existence of a large channel into gluons may give rise to a plethora of other particles, like anti-protons, and also opens interesting possibilities for direct detection of dark matter, as well as colliders signatures. 

In the sequel, we first derive model independent constraints from the observed anti-protons flux (section \ref{sec:anti-protons}). Then we consider more model dependent constraints that may be set by direct detection experiments, section \ref{sec:directdetection}. Finally we consider colliders constraints (jets+missing energy) and possible signatures in section \ref{sec:colliders}.

\section{Anti-proton flux}
\label{sec:anti-protons}

For DM particle with mass above the GeV scale the production of anti-protons from gluons is unavoidable
in the scenario we consider. Then, if we take for granted a positive gamma-ray line signal, an anti-proton flux is the most straightforward outcome, as the only relevant parameter is  the  ratio of Eq.(\ref{eq:ratio}).   More generally, given a scenario with both gamma and gluon lines, with a specific prediction for the ratio $C$, we may ask the following question: given the maximum anti-proton flux allowed, what is the prospect for gamma-ray line detection? In both cases, the constraints are independent of the topology considered. Also, independently of a $\gamma$ line, we may ask what is the maximum annihilation cross section into gluons allowed by the anti-proton data, and for which DM mass is it equal to $\langle \sigma v\rangle_{FO}$?

\begin{figure}[!htb]
\centering
\includegraphics[height=8cm]{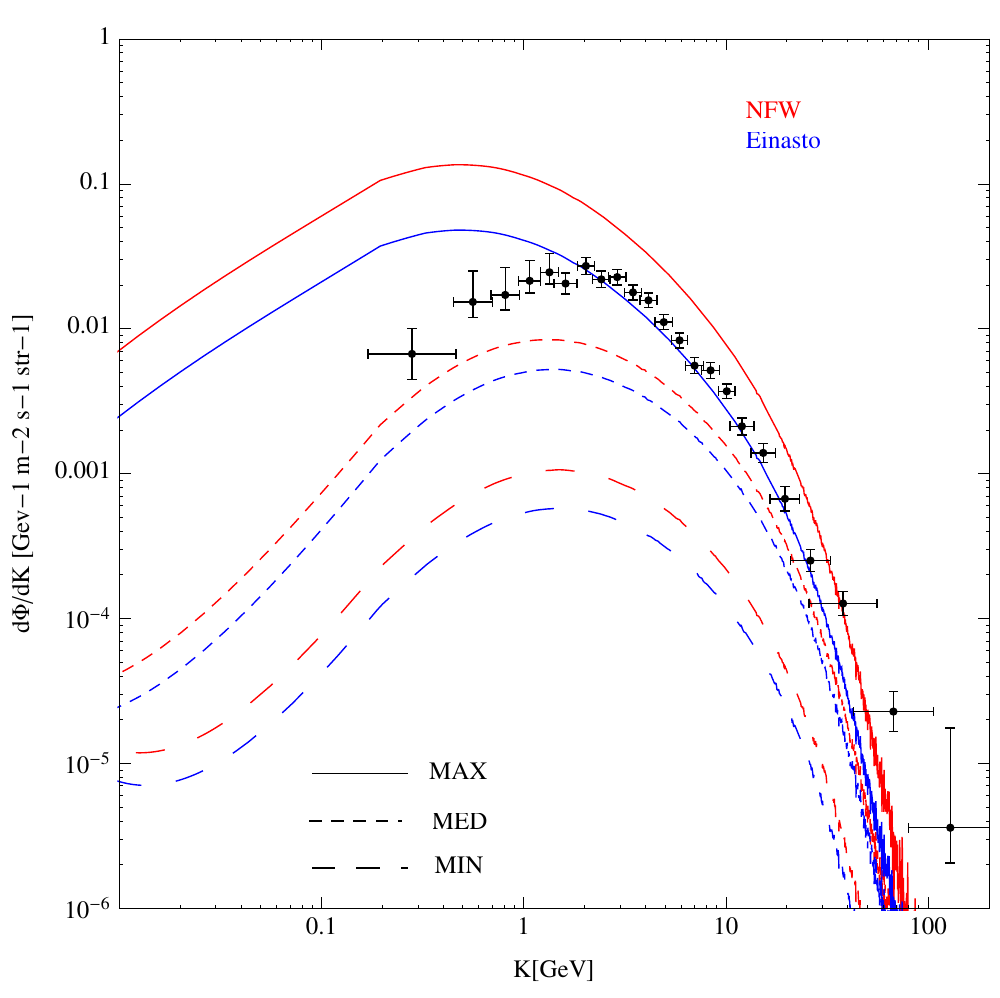}
\caption{Anti-proton fluxes from a WIMP candidate of $m_{DM} = 130$ GeV and annihilating in gluon pairs,  compared to the PAMELA measurements. The fluxes are given for the  NFW and Einasto profiles  and for MIN, MED and MAX cosmic propagation parameters.}
\label{fig:antipflux}
\end{figure}

\begin{figure}[!htb]
\centering
\includegraphics[height=5.5cm]{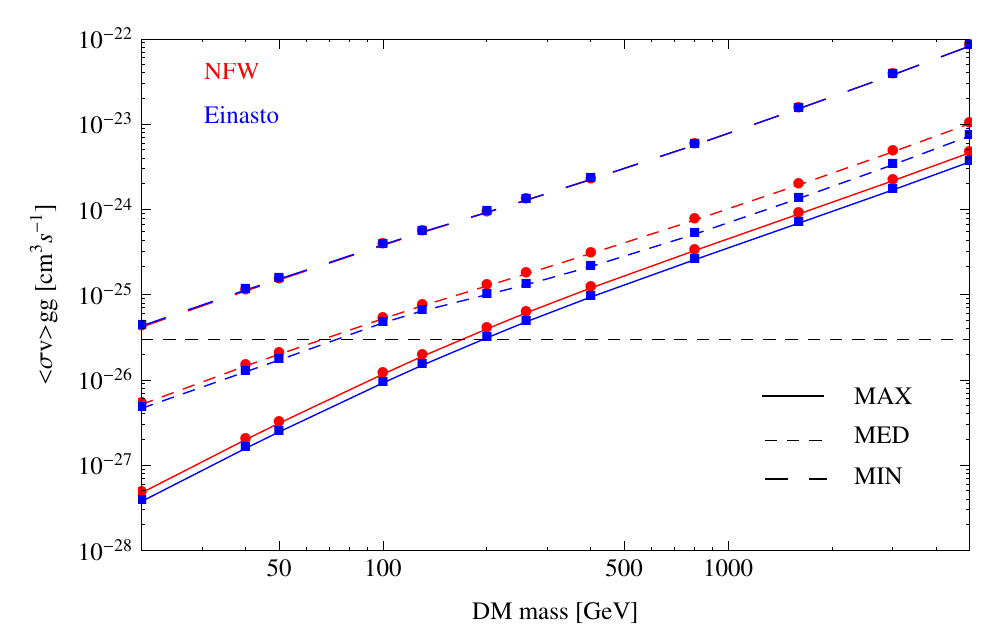}
\caption{Upper bounds on the annihilation cross section of DM into gluon pairs as set by the PAMELA anti-protons flux measurements for Einasto and NFW profiles, and three set of propagation parameters. The horizontal line corresponds to the standard thermal freeze-out annihilation cross section.}
\label{fig:maxGGantiP}
\end{figure}

\begin{figure*}
\includegraphics[height=11cm]{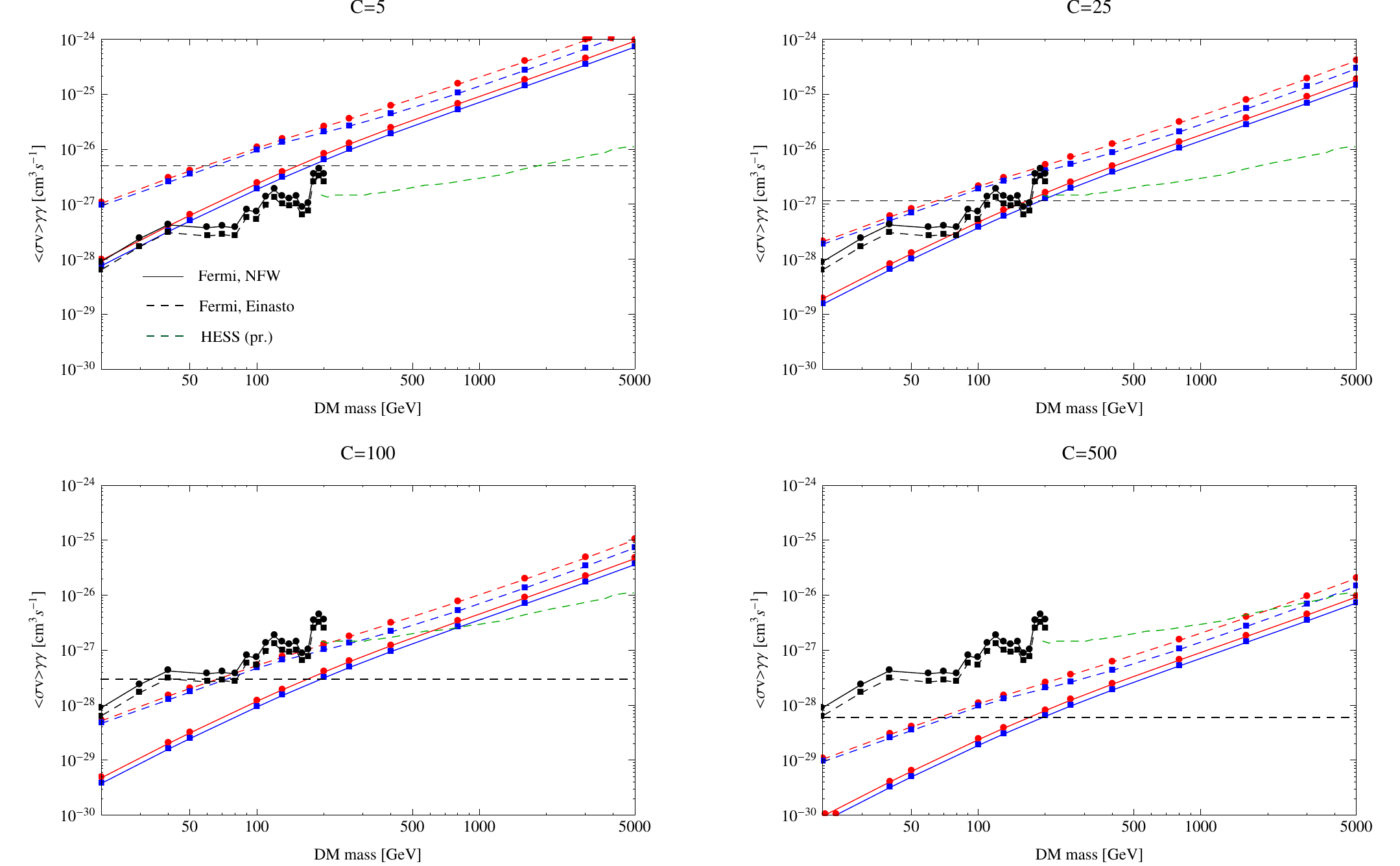}
\caption{From the results of Fig.~4, upper bounds on the DM annihilation cross section into monochromatic photons, for $C=5,\, 25,\, 100,\, 500$. The solid (dashed) black line gives the experimental upper bounds from Fermi-LAT assuming a NFW (Einasto) profile \cite{Ackermann:2012qk}. The green line is an estimate of the possible reach of the HESS experiment (taken from \cite{Bringmann:2011ye}). The horizontal line corresponds again to the FO annihilation cross section, rescaled by $C$.}
\label{fig:maxgammaflux}
\end{figure*}

In this section we refer specifically to bounds on annihilation into gluon and gamma-ray lines, hence to Eq.(\ref{ratiosigmaline}) \footnote{For secluded dark matter, each annihilation produces four photons with energy equal to $m_{DM}/2$. Hence, in Figs.(\ref{fig:maxGGantiP}) and (\ref{fig:maxgammaflux}), the horizontal line that corresponds to freeze-out abundance should be replaced by $\langle \sigma v\rangle =   \langle \sigma v\rangle_{FO}/2$, and the values  on the horizontal axis should be multiplied by $2$.}.
In Fig.\ref{fig:antipflux} we give, for illustration, the flux of anti-proton obtained assuming a single heavy vector quark of charge $Q_e=1$ and saturating the flux of gamma to that of the gamma line of Ref.\cite{Weniger:2012tx} at $130$ GeV.  In this plot, we have taken into account the contributions from both the $\gamma\gamma$ and $\gamma Z$ vertices. 
For the calculations of the flux of anti-protons we have used a popular semi-analytical approach, as summarized in Ref.\cite{Cirelli:2010xx}, to which we refer for more details. The figure displays two possible fluxes, using the standard  MIN, MED and MAX sets of cosmic ray propagation parameters and NFW and Einasto profiles for the distribution of dark matter in the Galaxy. The data points are those collected by the PAMELA collaboration \cite{Adriani:2010rc}. Basically, the conclusion is that the MIN and MED curves are consistent with the data, while  the flux corresponding to the MAX parameters overshoots the data and so is excluded.\footnote{A priori, another possible signal are positrons in cosmic rays, but we have checked that the flux of positrons is always substantially smaller than the PAMELA data. We have also checked that the overall flux of photons, including those from gluons, is consistent with various indirect constraints (diffuse galactic and extra-galactif gammas fluxes). }

Independently of the 130 GeV gamma-ray line, we give in Fig.\ref{fig:maxGGantiP} the bounds on  annihilation into {gluon} lines  allowed by the PAMELA anti-proton flux, taking into account the secondary anti-protons background flux produced by cosmic ray interactions \cite{Donato:2001ms}. 
Depending on the astrophysical setup, the bounds on annihilation cross-sections in gluon lines may be larger (or smaller) than the canonical freeze-out value. 
The standard way to read this plot is that, depending on the propagation parameters, a WIMP candidate annihilating dominantly into gluon lines, {\em i.e.} $\langle \sigma v\rangle_{gg} \simeq \langle \sigma v \rangle_{FO}$, is excluded below a given mass: $m_{DM} \sim 180$ GeV for MAX, $m_{DM} \sim 80$ GeV for MED, $m_{DM} \sim 16$ GeV for MIN (not shown). One should however keep in mind that  a larger cross section may be implemented in the Galaxy (at the prize of some fine-tuning and/or more model building) due to an astrophysical boost factor (however numerical simulations limits these to be of order a few \cite{Pieri:2009je}), or a particle physics effect (Sommerfeld enhancement, resonance,...). Our attitude here is agnostic, and we take these bounds as the maximum possible values allowed by current data (and the modelling of the CR background).

Now these bounds may be mapped on bounds on the maximum possible gamma-ray line fluxes. Although the mapping is straightforward, we find it useful to illustrate it in a separate plot. In  Fig.\ref{fig:maxgammaflux} we
show, for four possible values of $C = 5, 25, 100$ and $500$, 
the corresponding maximum annihilation cross sections in gamma ray lines from Fermi-LAT (at $95\%$ C.L.). In the same plot we show the estimation from Ref.\cite{Bringmann:2011ye} of the potential reach of the HESS experiment. 
Not surprisingly, a very low value, $C=5$, is basically excluded for a mass up to the value which has been probed by the Fermi-LAT experiment, $m_{DM}\simeq 200$~GeV, even for the MAX parameters.
At the other extreme 
with large values of $C$ (as exemplified for $C=500$), the expected flux are too small
to be probed by Fermi-LAT. 
The most interesting situation is the one of intermediate values, $C=25-100$.
For instance, for $C=100$ and MED propagation parameters, a WIMP, which emits gluons on top of photons with such a factor $C$, is  allowed by gamma lines for all masses.
The same situation holds for MAX parameters and $C=25$.
Alternatively adopting the MED parameters and $C\simeq25$, Fermi-LAT should have seen already a signal
with moderate significance, and the 130 GeV tentative line corresponds to such a case along our scenario.
Ultimately if a $\gamma$ line turns out to be accurately observed in the future, the results of Fig.~\ref{fig:maxgammaflux}
will allow to set an upper limit on $C$ (imposing that the annihilation cross section to gluons does not exceed the thermal freeze-out one). This will provide information on the ``beyond the SM'' heavy charged particles at the origin of this line. The discussion of this section also applies to narrow ``box-shaped'' spectrum by simply rescaling the mass of DM and the ratio $C$, according to relation (\ref{ratiosigmabox}).

\section{Direct detection}
\label{sec:directdetection}
A potentially interesting feature of the gluon line scenario is that it opens the possibility for direct detection through scattering on nuclei. The connection between the gluon cross section and possible direct detection is not as model independent as the anti-protons signature, but depends on the topology and on the DM particle nature. For some topologies the relation remains one-to-one, with, as we show,  prediction of a signal of the order the present experimental sensitivity.

\vspace{0.2cm}
\underline{Case 1. $\gamma$-line, quartic interactions}.
The most straightforward case is that of an effective quartic coupling between two DM particles and two gluons, which gives a one-to-one relation between the gluon cross section and the direct detection rate. In this case, a single scale, $\Lambda$, controls both the annihilation cross section and, for the gluon vertex, direct detection and signatures at hadron colliders, which we will discuss in section \ref{sec:colliders}, see Refs.\cite{Birkedal:2004xn,Bai:2010hh,Fox:2011pm,Goodman:2010ku,Cheung:2012gi}. 

For our argument, the most relevant operators involving two DM particles and two gluons are
\begin{eqnarray}
{\cal O}_{S}&=& \frac{1}{\Lambda_{S}^2}\chi_S^\dagger \chi_S\cdot\frac{\alpha_S}{12\pi} G^{a\mu\nu} G^a_{\mu\nu}
\label{OS}\\
{\cal O}_{F}&=& \frac{1}{\Lambda_{F}^3}\,i\bar{\chi}_F\gamma_5 \chi_F\cdot
\frac{\alpha_S}{8\pi} G^{a\mu\nu} \tilde{G}^a_{\mu\nu}
\label{OF}
\end{eqnarray}
for a scalar (S) and a fermion (F) dark matter candidate respectively (we consider complex scalar and Dirac fermion DM --- the generalization to real fields is straightforward). Generally speaking, one could also consider  fermionic operators involving $\bar{\chi}_F \chi_F$, but their annihilation is P-wave suppressed, and so they are irrelevant (baring a very large boost factor) for indirect detection, and {\em a forteriori} for a gamma ray lines signal. Similarly, one could think of operators mixing scalar and pseudo-scalar quantities, like 
$$
 \frac{1}{\Lambda_{S'}^2}\chi_S^\dagger \chi_S\cdot\frac{\alpha_S}{8\pi} G^{a\mu\nu} \tilde{G}^a_{\mu\nu}, \quad \frac{1}{\Lambda_{F'}^3}\,i\bar{\chi}_F\gamma_5 \chi_F\cdot \frac{\alpha_S}{12\pi} G^{a\mu\nu} {G}^a_{\mu\nu}.
$$
These operators are allowed but, as they break CP symmetry, they are in our opinion somewhat less likely  to have large coefficients, and for conciseness, we simply discard them. Hence we are left with the two possibilities of Eqs.(\ref{OS},\ref{OF}).

These operators lead to the following annihilation cross sections into gluon pairs: 
\begin{eqnarray}
\langle \sigma v\rangle_{gg} &=&\alpha_S^2 m_{\chi_S}^2/(9\pi^3 \Lambda_{S}^4)
\label{eq:OSsig}\\
\langle \sigma v\rangle_{gg} &=& \alpha_S^2 m_{\chi_F}^4/(2\pi^3 \Lambda_{F}^6)
\label{eq:OFsig}
\end{eqnarray}
Imposing a thermal freezeout abundance, Eq.~(\ref{sigmafo}), they give respectively
\begin{eqnarray}
m_{\chi_S} &\approx& 130 \;\left({\Lambda_{S}\over {134} \mbox{\rm GeV}}\right)^{2}\; \mbox{\rm GeV},\\
m_{\chi_F} &\approx& 130 \;\left({\Lambda_{F}\over 171 \mbox{\rm GeV}}\right)^{3/2}\; \mbox{\rm GeV},
\end{eqnarray}
were we set $\alpha_S = \alpha_S(130 \mbox{\rm GeV})$. 
\begin{figure}[!htb]
\centering
\includegraphics[height=8cm]{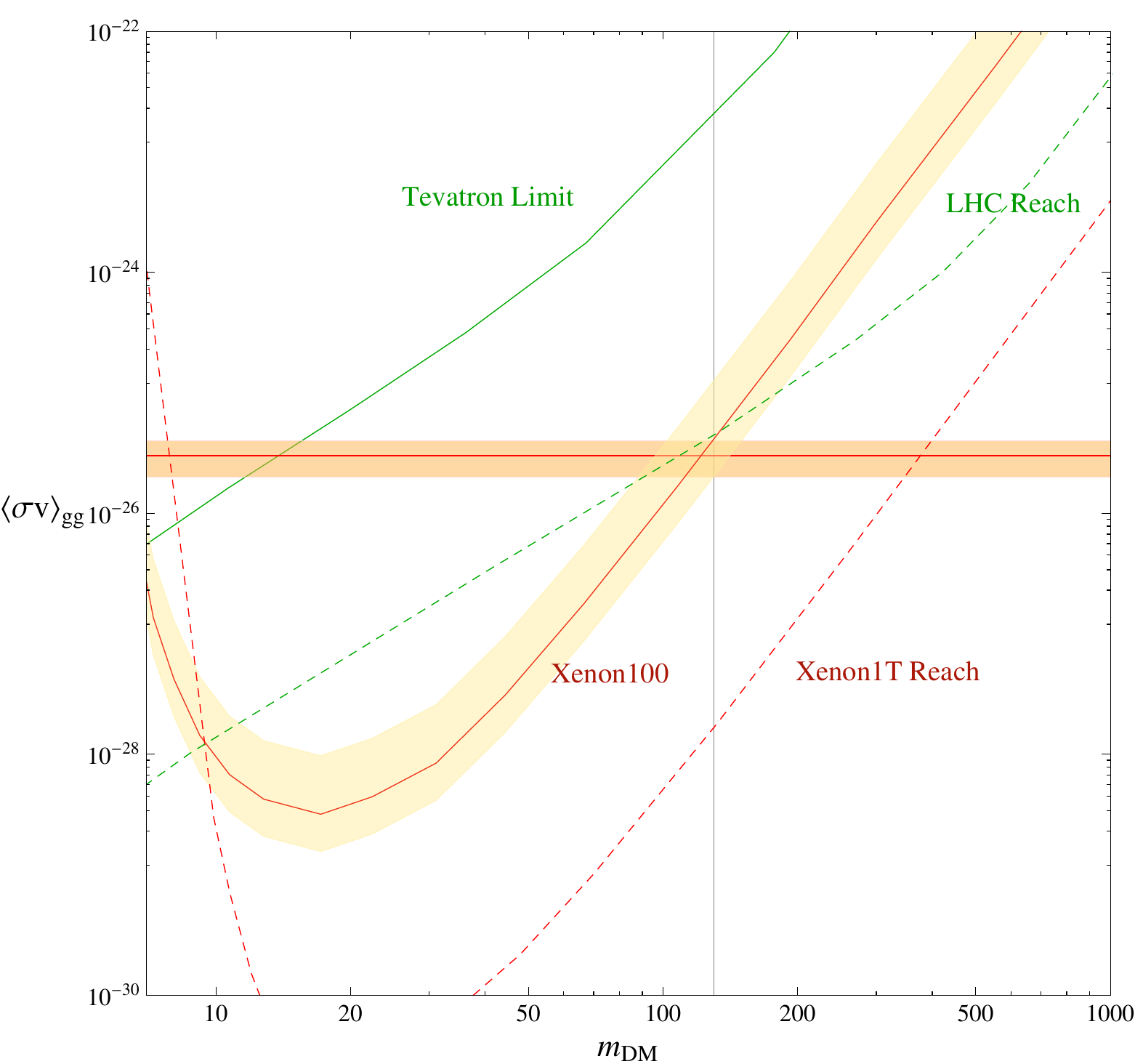}
\caption{Bounds from Xenon100 (solid) on a scalar candidate (masses in GeV) with coupling to gluons through the operator of Eq.(\ref{OS}), (annihilation cross section in cm$^3\cdot$s$^{-1}$), the Tevatron limits (solid) from jets + missing energy, and the prospects for Xenon1T (dashed, adapted from \cite{Aprile2010}) and the LHC (dashed, adapted from \cite{Goodman:2010ku}). For $\sigma v \sim 1$ pbarn (horizontal line), candidates heavier than $\sim 130$ GeV (vertical line) are allowed by current data.}
\label{fig:DDlimits}
\end{figure}

What are the prospects for direct detection (DD)? The operator (\ref{OF}) is {\em a priori}  out of reach of direct detection experiments because its scattering cross section is both recoil momentum suppressed, and Spin-Dependent (SD). The other operator however, Eq.(\ref{OS}), leads to a standard Spin Independent (SI) elastic cross section on nuclei, and so may be tested by DD experiments. The procedure to derive the effective coupling of $\chi_S$ to a nucleon is pretty straightforward \cite{Shifman:1978zn}. The matrix element of the gluon operator on a nucleon state may be written as
\begin{equation}
{\alpha_S \over 12 \pi} \langle N \vert G_{\mu\nu}^a G^{a \mu\nu}\vert N\rangle = f_N m_N \langle N \vert \bar \psi_N \psi_N\vert N\rangle (\equiv f_N m_N), 
\end{equation}
with an effective coupling $f_N = 2/27 \approx 0.074$. This is precisely the contribution that would be induced by a heavy quark loop. In our comparison with experiments, we use a more precise expression that takes into account the contribution from the light quarks to the nucleon mass (see for instance \cite{Bottino:2008mf}), 
$$
f_N = {2\over 27}\left(1 - ( 1 +  r y/2) \sigma_{\pi N}/m_N\right)
$$
where $y$ is the fractional strange-quark content of the nucleon, $r = 2 m_s/(m_u + m_d)$ and $\sigma_{\pi N}$ is the pion-nucleon sigma term. There are substantial uncertainties on $f_N$. Our estimate is $f_N = 0.053 \pm 0.011$ (at 1-$\sigma$). Regardless, from the above we get
\begin{equation}
\sigma_{SI}^{(N)} = {f_N^2\over 4 \pi} { m_N^2\, \mu^2\over \Lambda_{S}^4 \, m_{\chi_S}^2}
\label{eq:OSSI}
\end{equation}
where $\mu$ is the $\chi_S/N$ reduced mass. 

In Fig.\ref{fig:DDlimits}, we use Eqs.(\ref{eq:OSsig}) and (\ref{eq:OSSI}) to constrain the annihilation cross section into gluons using the exclusion limits set by Xenon100 \cite{Aprile:2011hi}. The horizontal band (reddish) corresponds to the standard freeze-out annihilation cross section {$\sim 1$} pbarn (see \cite{Steigman:2012nb} for a recent re-appraisal of $\langle\sigma v\rangle_{FO}$). The other band corresponds to the exclusion limits set by Xenon100. The width of the band corresponds to a 2-$\sigma$ variation of the parameter $f_N$. The vertical line corresponds to $m_{DM} = 130$ GeV. Also, we have allowed, in the relation between the annihilation cross section and the elastic SI scattering cross section, for the possibility of next-to-leading corrections in $\alpha_S$ to the annihilation cross section, which may amount to a factor $\lsim 2$, based on the expression for Higgs decay in {gluon-gluon} (see for instance \cite{Djouadi:2005gi}). The (corresponding) corrections to the elastic cross section are more difficult to assess, as the expression is non-perturbative in nature. To be concrete, we conclude that the Xenon100 data exclude all the candidates below $\sim 130$ GeV (except for light, $\sim$ few GeV, candidates, which we do not address here): hence a candidate with $m_{DM} =130$ GeV will be tested by the forthcoming Xenon100 data. The future Xenon1T experiment  may probe candidates up to  $\sim 385$ GeV. 

\vspace{0.2cm}
\underline{Case 2. $\gamma$-line, s channel exchange}. The operators of Eqs.~(\ref{OS}) and (\ref{OF}) can be induced by the s-channel exchange of, respectively, a scalar $S$ or  a pseudo-scalar $P$ particle (messenger in the sequel), Fig.\ref{fig:2gammatopologies}. For instance the following  Yukawa and one-loop effective interactions
\begin{eqnarray}
{\cal L} &{\supset}&  \mu_S S \chi_S^\dagger \chi_S + y_Q {S \over m_Q}\,{\alpha_S\over {12} \pi}G_{\mu\nu}^a G^{a \mu\nu}\,,
\label{eq:LagrS}\\
{\cal L} &{\supset}& i y_P P \bar{\chi}_F\gamma_5 \chi_F +  y_Q {P \over m_Q}\,{\alpha_S\over {8} \pi}G_{\mu\nu}^a \tilde G^{a \mu\nu}\,,
\label{eq:LagrF}
\end{eqnarray}
will induce such interactions. In the effective coupling between the messenger and gluons, we have in mind the effective coupling of a single heavy vector quark. More generally speaking, $m_Q/y_Q$ can be considered as being an effective mass scale, just like in operators (\ref{OS},\ref{OF}),  which encompasses the contributions of many degrees of freedom running in the loop. Accordingly, in the sequel  we will often set $y_Q=1$. 
The discussion of the relic abundance and of indirect signals depends on the mass of the s-channel
 messenger.

\vspace{0.14cm}
Clearly, for a \underline{heavy messenger}, $m_{S,P}^2>>4 m_{DM}^2$ 
, the discussion reduces to the previous case, Fig.~\ref{fig:DDlimits}. One recovers Eqs.~(\ref{OS})-(\ref{OF}) with
\begin{eqnarray}
{1\over \Lambda_{S}^2} &\equiv&   \;{y_Q\over m_Q} \;{\mu_S\over m_S^2}\,,
\label{eq:Seffective}\\
{1\over \Lambda_{F}^3} &\equiv&  \;{ y_Q \over m_Q} \;{y_P\over m_P^2}. 
\label{eq:Feffective}
\end{eqnarray}
This gives $\langle \sigma v\rangle_{gg}\simeq \langle \sigma v\rangle_{FO}$ provided
\begin{eqnarray}
m_{\chi_S}&=& 67\,\hbox{GeV}\cdot y_Q^{1\over 2} \cdot \Big(\frac{m_{\chi_S}}{m_Q}\Big)^{1\over2}\Big(\frac{\mu_S}{130 \,\hbox{GeV}}\Big)^{1\over2} \Big(\frac{4 m^2_{\chi_S}}{m_S^2}\Big)^{1\over2},
\nonumber\\
\label{eq:wmapscalS}\\
m_{\chi_F}&=&74\,\hbox{GeV}\cdot {y_Q} \cdot y_P\cdot \Big(\frac{m_{\chi_F}}{m_Q}\Big) \Big(\frac{4 m^2_{\chi_F}}{m_P^2}\Big).
\label{eq:wmapscalP}
\end{eqnarray}
Since $m_{DM}/{m_Q}$ and $4m^2_{DM}/m^2_{S,P}$ are smaller than one, candidates with $m_{DM}$ of order 130 GeV for example, require values of the couplings and of $\mu_S/m_{DM}$ that are much larger than one, typically  $\sim 10$. This condition is also required for the colour state $Q$ to be heavy enough to escape to the present collider bounds. Although these constraints could be slightly relaxed if there were several colored states, this discussion  illustrates how demanding it is to get a 130 GeV line from the operators of Eqs.~(\ref{OS})-(\ref{OF}) when rephrased in terms of an explicit model.

\vspace{0.14cm}
For the opposite case of a \underline{light messenger}, $4 m_{DM}^2> m^2_{S,P}$, one gets Eqs.~(\ref{eq:Seffective})-(\ref{eq:wmapscalP}) provided one substitutes $m^2_{S,P}$ by $4 m^2_{DM}$. In this case the last suppression factor of Eqs.~({\ref{eq:wmapscalS})-(\ref{eq:wmapscalP}) disappears, but still the condition
$\langle \sigma v\rangle_{gg}\simeq\langle \sigma v\rangle_{FO}$ requires quite large couplings.
The main difference with a heavy messenger is that  the elastic scattering on nuclei is  enhanced by a factor $(4m_{DM}^2/m_{S,P}^2)^2$. Therefore, given the bounds shown in Fig.~(\ref{fig:DDlimits}), such a scenario with  a candidate of  $m_{\chi_S}=130$~GeV and thermal relic abundance is excluded.
Interestingly, for the case of  a fermionic candidate, if $m_P$ is much lighter than $m_{\chi_F}$, one could get a large enough enhancement to have an observable spin-dependent signal, in spite of the  $q^4$ recoil momentum suppression. This enhancement is maximum for $m_{P}$ of order of the $\sim$~keV recoil energy probed in direct detection experiments. Concretely a ratio of $(m_P/m_{\chi_F})  \sim 10^{-4}$, which corresponds to $m_P \sim 10$ MeV for $m_{\chi_F} \sim 130$ GeV
would be required to be within reach of current SD direct detection experiments. 

\vspace{0.14cm}
Finally if $m_{S,P}\sim 2 m_{DM}$, we have two possibilities: 
\newline\noindent Firstly, we may have \underline{resonant} annihilation in the GC. While the bounds from direct detection (for a scalar candidate) are unchanged, this resonance boosts the signals in both gammas and gluons. These are  constrained by the bounds on the flux of anti-protons and gamma lines, see Figs.(\ref{fig:maxGGantiP},\ref{fig:maxgammaflux}). If annihilation in gluons fixes the relic abundance, then, for instance, the boost from the resonance is constrained to be less than $\sim 5$ (MED) for $m_{DM} = 130$ GeV. A corresponding bound on the gamma ray line flux may be read in Fig.(\ref{fig:maxgammaflux}): for the same setup, $C$ must be larger than $\sim 25$. 
For example an ${\cal O}(100)$ resonance boost of the $\gamma$ signal at 130 GeV would require a value of $C$ of order $\sim$ 2500. This is not inconceivable: for instance, the heavy fermion may carry color but have a small electric charge (take for example $Q_e=1/3$ in Eq.~(\ref{eq:goldenratio})); if there are more particles in the loops, then there is more freedom, including the possibility of playing with different mass scales.  
\newline\noindent Secondly, the resonance may be responsible for the relic abundance. Then both the direct and indirect signals are depleted, and the constraints from direct detection (again for the scalar DM) may be much weaker than those shown in Fig.~(\ref{fig:DDlimits}). In this case, if we focus on the 130 GeV line, and the relic abundance is dominantly from annihilation into gluons, then $C$ must be smaller than ${\cal O}(25)$, for otherwise $\langle \sigma v\rangle_{gg} \gsim \langle \sigma v \rangle_{FO}$. A way to accommodate a low $C$ value is to invoke, not only  heavy vector quarks, 
but also heavy vector charged leptons (possibly with smaller masses) \footnote{For a fixed value of  $\langle \sigma v\rangle_{gg} < \langle \sigma v \rangle_{FO}$, the constraints on the candidate from direct detection may be read (again for scalar DM) from Fig.(\ref{fig:DDlimits}).}. Of course, the limiting possibility is a scenario with no heavy quarks (and perhaps the relic abundance from resonant annihilation in $\gamma,Z$), in which case there are no constraints from direct detection. Such a scenario constitutes of course the extreme counterexample to the perspective we follow here. Still, in general one expects that $C\gg 1$  in beyond the SM scenarios.

\vspace{0.2cm}
\underline{Case 3. Box-shaped spectrum,  t-channel}. In case of secluded dark matter, the dark matter relic abundance and the indirect signal are not directly related. Consequently, if the discussion is the same as for $\gamma$ line for indirect signals,  the prospect for direct detection crucially depends on the $S$ particle lifetime.
Also, here  we only consider annihilation of a scalar particle, as annihilation of a fermionic candidate in the t-channel is $P-$~wave suppressed: this has little impact on the relic abundance, or direct detection, but it precludes indirect detection and so is of little direct interest for our purpose.

In the case of t-channel, the annihilation cross section of a scalar DM is given by 
\begin{equation}
\langle \sigma v\rangle_{t-channel} = \frac{\mu_S^4}{64 \pi m^2_{\chi_S}}\sqrt{1-\frac{m^2_S}{m_{\chi_S}^2}}\frac{1}{(2m^2_{\chi_S}-m^2_S)^2}\, , 
\end{equation}
where our conventions for the coupling between the scalar DM and the messenger $S$ are the same as in Eq.~(\ref{eq:LagrS}).
For reference, we give the decay rate of the $S$ particle in $\gamma\gamma$, $\gamma Z$  and gluons for the case of one heavy vector quark of charge $Q_e=1$:
\begin{equation}
\Gamma(S\rightarrow \gamma\gamma) = y_Q^2 {\alpha^2\over 16 \pi^3} {m_S^3\over m_Q^2},
\end{equation}
\begin{equation}
\Gamma(S \rightarrow \gamma Z) = 2 (1-m_Z/m_S)^3 \tan^2\theta_W \Gamma(S\rightarrow \gamma\gamma) 
\end{equation}
and
\begin{equation}
\Gamma(S\rightarrow g g)  = {y_Q^2 \alpha_S^2 m_S^3\over 72 \pi^3 m_Q^2}.
\end{equation}

The spectrum of photon (or gluons) produced per decay of the messenger has the shape of a box \cite{Ibarra:2012dw},
\begin{equation}
{dN\over dE} = {2 \over \Delta E} \Theta(E - E_{min}) \Theta(E_{max} - E)
\end{equation}
with $\Delta E = E_{max} - E_{min}$ and 
\begin{equation}
E_{max/min} = m_{\chi_S}/2 ( 1 \pm \sqrt{1 - m_S^2/m_{\chi_S}^2})\,. 
\end{equation}
 For this process, the WMAP relic abundance is obtained for \footnote{For the above processes to be relevant, we have to check that the annihilation in the s-channel of Fig.\ref{fig:2gammatopologies} (already discussed in the previous section) is sub-dominant. This will be the case provided $\mu_S \gsim  3  \left({\mbox{\rm TeV}\over m_Q}\right)$ GeV. A related  issue with 
DM annihilating  through a scalar messenger is that mixing with the Higgs must be small 
for  the dominant channel to be decay into gluon pairs. Concretely, for a $m_S \sim 260$ GeV messenger, we estimate that the bound on the S-Higgs mixing angle must be $
\theta \,\lsim\, 10^{-3} \left({\mbox{\rm TeV}\over m_Q}\right)
$.
As mixing is not prevented by any symmetry, such a small value is not natural. A 
pseudo-scalar would be better, as in this case one could invoke 
CP  to prevent mixing with the Higgs. We leave such considerations  for future work.} 
\begin{equation}
\mu_S \approx 160 \left({m_{\chi_S}\over 260 \mbox{\rm GeV}}\right)^{3/2} \, \mbox{\rm GeV}
\end{equation} 
for $m_{DM} \gg m_S$. If, on the contrary, the DM and the messenger are nearly degenerate, the box spectrum is narrow ({\em i.e.} close to  a line) and a good approximation for $\mu_S$, valid within $10\%$, is 
\begin{equation}
\mu_S \approx 152 \left({m_{\chi_S}\over 260 \mbox{\rm GeV}}\right)^{3/2} \left({0.1\over 
\epsilon}\right)^{1/8} \, \mbox{\rm GeV},
\label{wmapagain}
\end{equation} 
where  $\epsilon = (m_{DM} - m_S)/m_{DM}$. 

The direct detection cross section is that of Eq.(\ref{eq:OSSI}), with the scale $\Lambda$  given by Eq.(\ref{eq:Seffective}). A convenient way to write this cross section is  
\begin{equation}
\sigma_{SI} \approx 3.7 \cdot 10^{-41} \left({260 \mbox{\rm GeV}\over m_{\chi_S}}\right)^6  \left({\mu_S \over m_{\chi_S}}\right)^2 \left({\Gamma(S\rightarrow g g)\over m_{\chi_S}}\right)  \, \mbox{\rm cm}^2
\end{equation}
which is valid in the approximation $m_S \rightarrow m_{\chi_S}$. The appearance of the decay rate of the messenger (here in gluon pairs) is just a way to re-express the dependence in $y^2_Q/m_Q^2$. Since $m_S \approx m_{\chi_S}$, and $\mu_S$ may be fixed by the relic abundance, Eq.(\ref{wmapagain}), the cross section essentially depends on the DM mass, and the decay rate of the messenger (and also on the mass splitting, through Eq.(\ref{wmapagain})). To fix the ideas, we   consider 3 instances:\newline
\noindent
1) Take $m_{\chi_S} = 260$ GeV and $\mu_S = 180$ GeV (this corresponds to $\epsilon \approx 0.01$, $m_S = 257$ GeV). For this DM mass, the current exclusion limit set by Xenon100 is $\sigma_{SI} \lsim 2.2 \cdot 10^{-44}$ cm$^2$. This translates into an upper bound on $\Gamma(S\rightarrow g g ) \lsim 0.32$ GeV.\newline
\noindent
2) Take $m_{\chi_S} = 60$ GeV and $\mu_S = 20$ GeV ($\epsilon \approx 0.17$, $m_S = 50$ GeV). This candidate is interesting because it is for such mass that the Xenon100 constraints are the strongest. Taking  the current  Xenon100 limit  $\sigma_{SI} \lsim 6.8 \cdot 10^{-45}$ cm$^2$ we get that the decay rate has to satisfy $\Gamma(S\rightarrow g g )\lsim~1.5\cdot 10^{-5}$ GeV. 
\newline\noindent
3) Take $m_{\chi_S} = 130$ GeV and $\mu_S = 57$ GeV ($\epsilon \approx 0.04$, $m_S = 125$ GeV). This candidate is relevant for collider constraints (see next section). The Xenon100 limit is $\sigma_{SI} \lsim 1.2 \cdot 10^{-44}$ cm$^2$ so that $\Gamma(S\rightarrow g g )\lsim~3.2\cdot 10^{-3}$ GeV. 

Clearly, the bound from Xenon100 on the $m_{\chi_S} = 260$ GeV is not very strong (for $y_Q=1$ it would imply that $m_Q \gsim 18$ GeV)\footnote{This quantity should be considered as an effective mass scale, not necessarily the mass of a heavy quark.}. For the lighter candidate, thanks both to the stronger limit set by Xenon100 and  the scaling of the cross section with $m_{\chi_S}$ and the decay rate in $m_S$ and $m_Q$, the bounds on the scale $m_Q$ are quite relevant. For $y_Q=1$, $m_{\chi_S} = 60$ GeV gives $m_Q \gsim 300$ GeV, while for $m_{\chi_S} =130$ GeV, $m_Q \gsim 60$ GeV. For reference, the future Xenon1T experiment will bring these limits to:\newline
\noindent
1) $\Gamma (S \rightarrow gg ) \lsim 1.0 \cdot 10^{-3}$ GeV for $m_{DM}  =260$ GeV, corresponding to $m_Q \gsim 300$ GeV.
\newline\noindent
2) $\Gamma (S \rightarrow gg ) \lsim 7\cdot 10^{-8}$ GeV for $m_{DM} = 60$ GeV, corresponding to $m_Q \gsim 4.3$ TeV. The latter constraint is probably  stronger than the LHC reach. 
\newline\noindent
3)  $\Gamma (S \rightarrow gg ) \lsim 1.2\cdot 10^{-5}$ GeV for $m_{DM} = 130$ GeV, corresponding to $m_Q \gsim 1$ TeV. This is similar to the current LHC reach, see next section.

\vspace{0.2cm}
\underline{Case 4. Box-shaped spectrum, quartic coupling and  s-channel}.
We discuss these two cases only briefly, as the prospect for direct detection is either {\em null}, or it reduces to one of the previous cases. Indeed, if the messenger has a sizable vacuum expectation value, then we are back to a gamma-ray line through an s-channel (Case 2). If instead the vacuum expectation value is small or nonexistent, direct detection is doomed by the necessity of exchanging two messengers with the nucleon in an elastic scattering process, a process that is loop suppressed. A naive estimate gives (for the scalar candidate)
\begin{equation}
\sigma_{SI} \sim 3 \cdot 10^{-53} \left({1 \mbox{\rm GeV}\over m_S}\right)^2 \left({1 \mbox{\rm GeV}\over m_Q}\right)^4 \, \mbox{\rm cm}^2
\end{equation}
where we have only included the dominant parameter dependence (and have set $m_{DM} \sim 260$ GeV). 
This is true for the quartic coupling topology as well as for the s-channel one, with in the later case, perhaps, the  possibility of further suppression of the direct detection signal if a resonance occurs, as above. In principle one may have an inelastic process, with, say emission of a pair of photons, but we estimate this to be below the reach of direct detection experiments.


\section{Collider constraints}
\label{sec:colliders}

The predictions of our scenario at colliders depend, as for direct detection, on the specific topology of the relevant process. 

\vspace{0.2mm}
\underline{Case 1. $\gamma$-line quartic interactions}.
Again this is the  most straightforward case, as the interaction with gluons and/or photons depends only on one scale parameter. As in section \ref{sec:directdetection}, we consider the two operators of Eqs.(\ref{OS},\ref{OF}), which are the most relevant ones for indirect detection. Collider constraints on such operators rest on the possibility of creating DM pairs in collisions, the most obvious signature being the production of jets + missing energy. Taking into account the Standard Model background, limits may be set on the scale parameters, for each effective operator \cite{Birkedal:2004xn,Bai:2010hh,Fox:2011pm,Goodman:2010ku,Cheung:2012gi}.  Concretely, we  refer to the analysis of Ref.\cite{Goodman:2010ku}, where the collider constraints have been studied.\footnote{The operators of Eqs.(\ref{OS},\ref{OF}) correspond respectively to the operators $C_5$ and $D_{14}$ in \cite{Goodman:2010ku}.} 
In Fig.\ref{fig:DDlimits}, we have reported the bounds from  \cite{Goodman:2010ku}, both from the Tevatron and the prospect for the LHC ({\em i.e.} for $E_{CM} = 14$ TeV, and $30$ fbarn$^{-1}$). The bottom line is that the current collider limits are less constraining (but are independent) from those from  direct detection searches.\footnote{The limits from the recent LHC runs are not available for the effective operators of DM into gluons, Eqs.(\ref{OS},\ref{OF}). The current LHC constraints on other operators, like effective couplings to quark-anti-quark pairs, give bounds that are only a factor of a few stronger than those based on Tevatron data, and weaker  than the current Xenon100 bounds (for Eq.(\ref{OS})). The same conclusion, we expect, should hold for the gluonic operators. } On the long run, the LHC may exclude candidates  $m_{\chi_S} \lsim 100$ GeV, but these will be superseded by Xenon1T (if no DM events are observed).  Similar, but slightly stronger conclusions may be reached for the fermionic candidate, see Fig.(11) in Ref.\cite{Goodman:2010ku} for which, we recall, there are no limits from direct detection: the Tevatron data exclude a fermionic candidate below $m_{\chi_F} \lsim 40$ GeV; the prospect for the LHC is $m_{\chi_F} \lsim 700$ GeV, assuming the standard freeze-out abundance.

\vspace{0.2mm}
\underline{Case 2. $\gamma$-line, s-channel exchange}. Regarding the jets + missing energy constraints, the discussion for the second diagram of Fig.~1 reduces to the previous case if the mass of the s-channel messenger particle is heavier than the other energy scales, $m_{S,P}^2 \gg 4 m_{DM}^2$ and $m_{S,P}^2\gg s$ where $s$ refers to the momentum square of the particle $P$ at LHC. If instead $m^2_{S,P}$ is light enough to be produced at Tevatron or LHC, with $4m_{DM}^2< m_{S,P}^2$, the production is resonantly enhanced which strengthens the collider sensitivity (whereas the annihilation cross section is not enhanced). If on the contrary $4m_{DM}^2\sim m_P^2$,  the annihilation cross section is resonantly enhanced, in which case the collider constraints are weaker than those for the quartic interaction (Case 1).

This brings  another possible signature at colliders, which is the production at the LHC of an on-shell messenger particle through gluon-gluon fusion, followed by its decay into a pair of gammas. We discuss the prospect for this signal in Case 3, for which such a signal is a natural prediction.

\vspace{0.2mm}
\underline{Case 3. Box-shaped spectrum, t-channel}. Regarding the jets + missing energy constraints, the same comments as in Case 2 apply, so we do not repeat them here. 

If we focus on the possibility of a narrow box-shaped spectrum, which implies that $m_{DM} \approx m_{S,P}$, then one prediction of this scenario is the on-shell production of the messenger particle at colliders. The process is precisely similar to the production of the Higgs, so we will heavily rest on what is known for this particle. The smoking signal is the production of di-photons, with invariant mass $\sim m_{DM}$ (a clean signal). The production of a pair of gluon jets is also potentially interesting, but the background is much larger. 

The gluon-gluon to gamma-gamma ratio of production cross section is directly related to the ratio of Eq.(\ref{eq:ratio}). For the sake of illustration, we will consider two DM candidates with $m_{DM} = 260$ GeV and $m_{DM} = 130$ GeV.  As before, we consider the heavy fermion limit ($4m_Q^2 \gsim m_S^2$), in which case
 the following relation is  a good approximation,
\begin{equation}
\frac{\sigma (pp(gg)\rightarrow S+X)}{\sigma(pp(gg)
    \rightarrow H+X)}\approx \frac{\sigma_{LO}(gg \rightarrow S)}{\sigma_{LO}(gg
      \rightarrow H)} \approx 0.045 \,\left(y_Q {\mbox{\rm TeV}\over m_Q}\right)^2,
\end{equation}
where  $\sigma (pp(gg)\rightarrow H+X)$ is the Higgs
boson production cross section through gluon-gluon fusion. Taking the latter for $m_H = 260$ GeV, we get from Ref.\cite{Dittmaier:2011ti}
\begin{eqnarray}
\sigma (pp(gg)\rightarrow H+X)&=&\text{3.12 (13.43) Pb at $\sqrt s$=7 (14) TeV},\nonumber\\
\end{eqnarray}
from which we infer that the production cross section of a $m_S \simeq 260$ GeV particle $S$
is
\begin{equation}
\sigma (pp(gg)\rightarrow S+X)= 0.14\, (0.61) \left(y_Q {\mbox{\rm TeV}\over m_Q}\right)^2\,
\mbox{\rm pb}\label{scalar:Pb}
\end{equation}
 at $\sqrt s =7\, (14) \mbox{\rm TeV}$. The branching ratios of Higgs in gluon pairs or di-photons at $m_H = 260$ GeV are small \cite{Dittmaier:2011ti}:
\begin{eqnarray}
Br(H\rightarrow \gamma \gamma) &\approx & 2 \times 10^{-5},\\
Br(H\rightarrow gg)&\approx&7\times 10^{-4},\nonumber
\end{eqnarray}
while the $S$ decays dominantly into gluon pairs. (We note in passing that the ratio of these branchings is $C_H \approx 32$.) From the above numbers, we infer that 
the number of di-photons events from $S$, normalized to that from a Higgs at $m_H = 260$ GeV, is
\begin{equation}
\left. {N(S\rightarrow \gamma \gamma)\over N(H\rightarrow \gamma \gamma)} \right\vert_{m_S = 260 GeV} \approx {2 \cdot 10^3\over C} \times \left(y_Q {\mbox{\rm TeV}\over m_Q}\right)^2.
\end{equation}
We did not find constraints on di-photons  in the literature (although they must exist) for such a large invariant mass scale. For comparison, let us consider the sensitivity one may have for  $m_S \sim 125 $ GeV (corresponding to $m_{DM} \sim 130$ GeV, see Section \ref{sec:directdetection}). In this case we may refer to the current constraints on di-photons from Higgs searches at the LHC \cite{ATLAS:2012ad}.  For $m_H = 125$ GeV, we have
\begin{equation}
\left.{N(S\rightarrow \gamma \gamma)\over N(H\rightarrow \gamma \gamma)} \right\vert_{m_S = 125 GeV}\approx {30\over C} \times \left(y_Q {\mbox{\rm TeV}\over m_Q}\right)^2.
\label{eq:ratio125}
\end{equation}
An excess of di-photons at such energy is quite constrained by the current LHC data, in particular if we assume that the Higgs mass is indeed $125$ GeV (see for instance Ref.\cite{Carmi:2012yp}). If we assume that the ratio of Eq.(\ref{eq:ratio125}) is constrained to be $\lsim 1$ by LHC data,  we have $m_Q \gsim 1$ TeV for $y_Q= 1$ and $C \sim 30$. This is stronger than the current Xenon100 bound (see Section \ref{sec:directdetection}): $m_Q \gsim 60$ GeV (for the same Yukawa coupling) and comparable to our estimate of the Xenon1T sensitivity  reach,  $m_Q \gsim 1$ TeV. We tentatively conclude that the constraints on such processes are in a ballpark at the interface of direct detection and colliders, and are worth being investigated more systematically, something we leave for future work. 

\vspace{0.2cm}
\underline{Case 4. Box-shaped spectrum, quartic coupling and  s-channel}.
  If the messenger has a sizable vacuum expectation value, then the situation regarding the constraints from jets + missing energy is analogous to that of Cases 2 and 3. 

If not, these topologies imply that jets + missing energy are accompanied by the emission of gluons (or photon) pairs from the second messenger particle in the diagram. Assuming that the latter is off-shell (we do not consider the possibility of two different messenger mass scales) we expect that these processes are much suppressed compared to the signals discussed in Cases 2 and 3, and so are not very constraining.

Regardless, we may also create the messenger particle and observe its decay into di-photons, as discussed in the previous section (Case 3).

\section{Summary and Prospects}

The observation of a sharp $\gamma$ feature at the level of sensitivity of the Fermi-LAT (or future Cerenkov telescope arrays) could be explained from  annihilation of dark matter particles. Such a process is expected to occur through loop diagrams, and may be associated to   various topologies, which at one-loop are shown in Figs.\ref{fig:2gammatopologies},\ref{fig:boxtopologies}. 
Such a signal would obviously point to a relatively large annihilation cross section into $\gamma$, which could be explained  either through large couplings and/or the proximity of a resonance (gamma-line) or, without the need of any of such features, through a secluded dark matter candidate  (box-shaped spectrum).
A natural question one may ask is: what if the heavy charged particles in the loop, which emit the photons, also carry color charge? After all this is the case for the classic instances: $\pi_0$, Higgs and axions. Annihilation in gluon pairs is allowed in many dark matter models (in particular the MSSM), but suppressed. 
Here we consider the simple possibility that the annihilation is mostly into gluons. This may arise in a framework in which the dark matter particle (and its siblings) lives in a hidden sector, and  interacts with the visible sector only through heavier particles. In this work we have studied several aspects of  such a scenario in a model independent way (as far as possible). In particular,
Fermi-LAT is currently probing annihilations into  gamma ray lines that are 1 or 2 orders of magnitude below the freeze-out cross section. If a line is
 observed in a near future, then the hypothesis of annihilation into gluon pairs could naturally explain the observed relic dark matter density. This holds in particular for the tentative 130 ~GeV $\gamma$ line claimed in Ref.~\cite{Weniger:2012tx}. 

If DM annihilates dominantly into gluon pairs, an inevitable by-product is a large flux of anti-protons, which is constrained by PAMELA. 
The upper bounds on the annihilation cross section of DM into gluon pairs depend only on the propagation setup, and not on  specific particle model considerations. Using a standard semi-analytical approach, we give in Fig.\ref{fig:maxGGantiP} the bounds on $\langle \sigma v\rangle_{gg}$ for three sets of CR propagation parameters (MIN, MED and MAX) and two DM profiles (NFW and Einasto, but the dependence on the profile is mild). The bounds depend sensitively on the choice of propagation parameters: a thermal relic candidate is allowed for  $m_{DM}\gsim 180$ GeV in the MAX setup,  $m_{DM} \gsim 80$ GeV for MED, and $m_{DM} \gsim 16$ GeV for MIN. More generally, and independently of the relation with thermal freeze-out,  but assuming specific values for the gluon-to-gamma emission ratios, Eq.(\ref{eq:ratio}), the anti-protons data set upper bounds on the annihilation cross section into $\gamma$, see Fig.~5.

Annihilation into gluons also opens the possibility for direct detection. The connection between direct detection rates and the gluon/photon cross sections is more model dependent, as it depends on the topology of the process and the nature of the DM candidate. For  $\gamma$-ray lines through a quartic interaction, the direct detection rate is in one-to-one correspondence with $\langle \sigma v\rangle_{gg}$. For a scalar DM candidate with WMAP abundance, Xenon100 excludes candidates below  $m_{DM} \lsim  130$~GeV, see Fig.\ref{fig:DDlimits}. In particular, the predicted rate for the 130 GeV line is at the edge of the current exclusion limits. For the other topologies, the connection is looser, and they may lead to observable direct detection signals, see Section \ref{sec:directdetection}. 

A DM candidate that couples to gluons is constrained by collider searches, through jets + missing energy signals. Again, the correspondence  is most straightforward for quartic effective operators. Using the results of  Ref.\cite{Goodman:2010ku}, we put in perspective the current and forthcoming constraints from colliders. As far as we know, the impact of the current data on the operators of Eqs.(\ref{OS},\ref{OF}) has not yet been shown by the LHC experiments. It would be interesting to do so. Even if the current bounds are weaker than the current direct detection exclusion limits for a scalar candidate, they provide independent constraints. In the case of the fermionic candidate, collider data should set the best current constraints. As for the other topologies, we have discussed to which extent they relate to the constraints on the quartic interactions. Finally, we have put forward the possible production of the intermediate scalar
 (or pseudo-scalar) particle through gluon-gluon fusion, which may decay into two photons, a process reminiscent of (light) Higgs searches. 

Our results show that it could be interesting to further constrain such a  process using  available and forthcoming LHC data. One last potentially interesting signal (work in progress) is the capture of DM in the Sun, and its annihilation into a hard spectrum of neutrinos. As emphasized by many authors, annihilation into a pair of photons should also annihilate into $\gamma Z$. The latter may decay into a pair of neutrinos, with energy $E_\nu \sim m_{DM}/2$ (for a $Z$-line), or $E_\nu \sim m_{DM}/4$ (for a narrow box-shaped $Z$ spectrum). 


\vspace{5mm}
\section*{Acknowledgement}
\vspace{-3mm}
One of us (M.T.) acknowledges stimulating discussions with Y. Mambrini and U. Ellwanger, and the LPT-Orsay for hospitality. 
 T.H. thanks the Departamento de F\'isica Te\'orica (UAM-Madrid) and the IFT-Madrid for hospitality and the Comunidad de Madrid (Proyecto HEPHACOS S2009/ESP-1473). This work is supported by the FNRS, the IISN and  an ULB-ARC.

\end{document}